\documentclass[5p]{elsarticle}

\usepackage[T1]{fontenc}
\usepackage{setspace}
\usepackage{graphics}
\usepackage{amsmath}
\usepackage{textcomp}
\usepackage{array,booktabs}

\begin{document}

\begin{frontmatter}

\title{Reverse Monte Carlo investigations concerning recent isotopic substitution neutron diffraction data on liquid water}
\author{Ildik\'o Pethes}
\cortext[mycorrespondingauthor]{Corresponding author}
\ead{pethes.ildiko@wigner.mta.hu}
\author{L\'aszl\'o Pusztai}
\address{Wigner Research Centre for Physics, Hungarian Academy of Sciences, Konkoly-Thege M. \'ut 29-33, 1121, Budapest, Hungary}

\begin{abstract}

Although liquid water has been studied for many decades by (X-ray and neutron) diffraction measurements, new experimental results keep appearing, virtually every year. The reason for this is that neither X-ray, nor neutron diffraction data are trivial to correct and interpret for this essential substance. Since X-rays are somewhat insensitive to hydrogen, neutron diffraction with (most frequently, H/D) isotopic substitution is vital for investigating the most important feature in water: hydrogen bonding. Here, the two very recent sets of neutron diffraction data are considered, both exploiting the contrast between light and heavy hydrogen, $^1$H and $^2$H, in different ways. Reverse Monte Carlo structural modeling is applied for constructing large structural models that are as consistent as possible with all experimental information, both in real and reciprocal space. The method has also proven to be useful for revealing where possible small inconsistencies appear during primary data processing: for one neutron data set, it is the molecular geometry that may not be maintained within reasonable limits, whereas for the other set, it is one of the (composite) radial distribution functions that cannot be modeled at the same (high) level as the other three functions. Nevertheless, details of the local structure around the hydrogen bonds appear very much the same for both data sets: the most probable hydrogen bond angle is straight, and the nearest oxygen neighbours of a central oxygen atom occupy approximately tetrahedral positions.

\end{abstract}


\end{frontmatter}

\section{Introduction}

Liquid water, as it is the the basis of life on Earth, is a most common, everyday liquid. It is the simplest compound of two universal elements, hydrogen and oxygen ($\mathrm{H_2O}$), and is the second most frequent molecule in the Universe. In spite of these, water is still an amazing substance with unique properties: it has phase, density, thermodynamic and other physical anomalies \cite{MartinChaplinWebsite}. Hence it is not surprising that water has been, and still is the most researched liquid. (For a comprehensive summary, see e.g. the collection of references in \cite{MartinChaplinWebsite}.)

In order to understand its properties (and anomalies!), it is necessary to comprehend the microscopic structure of liquid water. There are numerous studies in the literature using different spectroscopy and scattering techniques: X-ray \cite{Hura2000, Sorenson2000, Skinner2013} and neutron scattering \cite{Soper2000, Zeidler-Salmon2011, Zeidler-Salmon2012, Soper2014}, X-ray absorption spectroscopy (XAS) \cite{WernetScience2004, Myneni2002, Naslund2005, Smith2004}, X-ray emission spectroscopy (XES) \cite{Tokushima2008, Fuchs2008, Odelius2005, Forsberg2009} or small-angle X-ray scattering experiments (SAXS) \cite{Huang2009, Clark2010PNAS}. To interpret the experimental results many models and computer simulation works have been performed (see e.g. \cite{Soper2000,Pusztai1999,Guillot2002,Clark2010, Paesani2009} and references therein).

Though there have been many investigations, there is not a reassuring consensus present about the structure of water. For example, even the concept of tetrahedral arrangement of 4-fold coordinated water molecules, that had been widely accepted earlier,  has been challenged and an asymmetric distribution of twofold-coordinated molecules was suggested \cite{WernetScience2004}. The proposed structural organization, with hydrogen-bonded chains and rings of water molecules in a weakly hydrogen-bonded disordered network, is still controversial (see e.g. \cite{Soper2005JPCM, Leetmaa2008}). As an interesting addition to the ongoing debate concerning coordination numbers, Skinner et al. report an 'isosbestic point', a distance at which the integrated O-O coordination number is independent of temperature\cite{Skinner2014}.

Diffraction measurements are the obvious means for structure determination. X-ray diffraction can be useful for determining the oxygen-oxygen (and, to some extent, oxygen-hydrogen) correlations\cite{Skinner2014}, but this technique is not too sensitive for hydrogen. Since neutron scattering lengths are different for the hydrogen isotopes, neutron diffraction with H/D isotope substitution is, in principle, suitable for the separation of the hydrogen-hydrogen and hydrogen-oxygen correlations (if one accepts the approximation that the structure of water depends only weekly on the isotopic composition). In practice, however, the correction of the measured scattering data is extremely difficult due to the strong incoherent and inelastic scattering of neutrons by protons ($^1$H). There is still not a generally applicable way to remove such effects from the measured data (see, e.g., \cite{Soper2000,Pusztai1999,Temleitner2007}). This is why structural modeling \cite{McGreevyPusztai1988,Soper2005PRB} has played an important role \cite{Soper2000,Soper2014,Pusztai1999} in the history of structure studies on liquid water.

One of the possibilities that can be useful for interpreting the corrected diffraction data sets is to prepare three-dimensional atomic (structural) models that are consistent with all of the input data sets. The Reverse Monte Carlo (RMC) method \cite{McGreevyPusztai1988} is a perfect tool for this purpose. RMC also allows one to check whether it is possible to prepare physically meaningful particle configurations that match corrected experimental data (see Ref.\cite{Gereben1994b}); i.e., the method (under certain conditions) may serve for providing a kind of 'first aid support' to measured diffraction data.

In this work we have studied two sets of very recent neutron total scattering structure factors (TSSF) and the corresponding radial distribution functions. One of them \cite{Soper2014}, referred to as 'Case 1' throughout, presented diffraction measurements on four mixtures of light and heavy water; the partial radial distribution functions (PRDF) were also provided. The other study \cite{Zeidler-Salmon2012}, indicated as 'Case 2' in the following, is based on neutron diffraction measurements with oxygen isotope substitution. From these measurements the authors obtained linear combinations of the O-O and O-H (or O-D) partial structure factors. They presented radial distribution functions obtained by Fourier transformations, as well. 

Two important differences between the two sets of neutron diffraction experiments should be mentioned explicitely: 

(1) Data for 'Case 1' have been gathered at a spallation neutron source (ISIS, UK), whereas experiments connected to 'Case 2' were performed on a steady-state reactor based instrument (ILL, France); this difference has a profound effect on the methods and complexity of data treatment (see Refs.\cite{Soper2014,Zeidler-Salmon2012}).

(2) Since it was not stated otherwise in Ref.\cite{Soper2014}, standard values of the atomic coherent scattering lengths were assumed for data in 'Case 1', whereas for 'Case 2', they were taken from the original publication\cite{Zeidler-Salmon2012}. The coherent scattering lengths applied throughout this work are quoted in Table \ref{tab:scattering_lengths}.

\begin{table}[ht]
  \centering
  \renewcommand{\arraystretch}{1.5}
   \begin{tabular}{>{\centering\boldmath}m{0.16\columnwidth} >{\centering}m{0.14\columnwidth} >{\centering}m{0.14\columnwidth}>{\centering}m{0.14\columnwidth}>{\centering\arraybackslash}m{0.14\columnwidth}}
    \toprule
    & \boldmath{$b_{\mathrm{{ }^{nat}O}}$}& \boldmath{$b_{\mathrm{{ }^{18}O}}$}&\boldmath{$b_{\mathrm{D}}$} & \boldmath{$b_{\mathrm{H}}$}\\
    \midrule
    Case 1 &5.803 &- & 6.671 & -3.7406\\
    Case 2& 5.805& 6.005 & 6.619 & -3.7409\\
    \bottomrule
  \end{tabular}
  \caption{\label{tab:scattering_lengths} Coherent atomic scattering lengths used for neutron data of 'Case 1'\cite{Soper2014} and 'Case 2'\cite{Zeidler-Salmon2012} (in fm).}
\end{table}

We performed Reverse Monte Carlo modeling, in order to (a) learn about the internal consistency of these new, much improved scattering data and (b) find structural models that would be consistent with the entire data sets, including the $r$-space information, simultaneously.

\section{Reverse Monte Carlo modeling}

Reverse Monte Carlo (RMC) modeling has been described in many publications in detail \cite{McGreevy2001,Evrard2005,Gereben2007,Gereben2012}, so here only a brief description is necessary. RMC is an inverse method to obtain large three-dimensional structural models that are consistent with the supplied (experimental and/or theoretical) data sets (TSSF-s and/or (P)RDF-s). It can be used in conjunction with any quantity that can be expressed directly from the atomic coordinates. By moving particles randomly in the simulation box, the difference between the experimental (or 'quasi-experimental', see, e.g., \cite{Pusztai2008a} and RMC model structural quantities (e.g. structure factors) is minimized. As a result, particle configurations are obtained that are consistent with all the input data. From the particle configurations, further structural characteristics (coordination numbers, nearest neighbor distances, bond angle distributions, etc...) may be calculated. Over the past nearly three decades, RMC has been successfully applied to a wide variety of systems, from simple liquids (see, e.g., the review of McGreevy \cite{McGreevy2001}), through metallic \cite{Jovari2007} and covalent \cite{Jovari2008} glasses, simple \cite{Pothoczki2012,Temleitner2014} and hydrogen-bonded \cite{Pusztai1999} molecular liquids, as well as for characterizing disordered crystalline structures \cite{Tucker2007,Pothoczki2013}.

The method is also suitable for establishing whether various input data sets are consistent with each other: if they are then they can be fitted simultaneously within their uncertainties \cite{Pusztai2008a}. If, on the other hand, not each element of the input set is consistent with the others then it is possible to tell which element is problematic: an approach proposed and tested recently \cite{Steinczinger2012} for input data consisting of TSSF-s and PRDF-s is applied in this work.

Here the RMC++ code\cite{Gereben2007} was used to obtain structural models. The cubic simulation boxes contained 6000 atoms (2000 molecules), the atomic number density was 0.1 \AA$^{-1}$, the simulation box length was 39.1 \AA. In order to keep the atoms together in the molecules during the calculations the 'fixed neighbor constraints' (FNC) option was applied \cite{Evrard2005}. This algorithm connects two hydrogen atoms and their oxygen central atom permanently via their identity numbers (defined for the atomic configuration file). For the realization of this constraint, the O-H and H-H intramolecular distances were kept between minimum and maximum values: the values for the various FNC combinations tested are shown in Table \ref{tab:fnc}. Intermolecular closest approach ('cutoff') distances between atoms were chosen as indicated in Table \ref{tab:fnc}.

\begin{table}[ht]
  \centering
  \renewcommand{\arraystretch}{1.5}
  \begin{tabular}{>{\centering\bfseries}m{0.35\columnwidth} >{\centering}m{0.14\columnwidth} >{\centering}m{0.14\columnwidth} >{\centering\arraybackslash}m{0.14\columnwidth}}
    \toprule
     & \textbf{O-O} & \textbf{O-H} & \textbf{H-H} \\
    \midrule
    intramolecular distances realistic molecule & -- & 0.95 - 1.02 & 1.50 - 1.62 \\
    intramolecular distances 'elastic' molecule & -- & 0.78 - 1.15 & 1.36 - 1.71\\
    intermolecular distances \\all models& 2.3 & 1.5 & 1.6\\
    \bottomrule
  \end{tabular}
  \caption{\label{tab:fnc} Minimum and maximum O-H and H-H intramolecular distances used in fixed neighbor constraints and intermolecular closest approach distances (cutoff distances) between atoms (in \AA{}).}
\end{table}

Simulations were started from a random configuration where only the cutoff distances and the fixed neighbor constraints were in effect ('hard sphere' model). The number of accepted moves was about $1-2 \times 10^7$ in each calculation. $\sigma$ parameters (essentially, control parameters for the different data sets that influence the tightness-of-fit, cf. Ref.\cite{Evrard2005}) were decreased progressively during the simulation runs, resulting in a gradually improving fit to the target functions. Final $\sigma$ values are listed in Tables \ref{tab:sigmaSoper} and \ref{tab:sigmaSalmon}. Note that these tables also provide information on the sets of input data applied in the various RMC calculations: for Case 1 (data from Ref. \cite{Soper2014}), always at least 7 target functions (4 TSSF-s, 3 intermolecular PRDF-s) were considered, whereas for Case 2 (data from Ref. \cite{Zeidler-Salmon2012}), 4 functions (two in $Q$ and two in $r$ space, each a kind of 'composite') were approached.

\begin{table}[ht]
  \centering
  \renewcommand{\arraystretch}{1.5}
  \begin{tabular}{>{\centering\boldmath}m{0.19\columnwidth} >{\centering}m{0.14\columnwidth} >{\centering}m{0.14\columnwidth}>{\centering}m{0.14\columnwidth}>{\centering\arraybackslash}m{0.14\columnwidth}}
    \toprule
    & \boldmath$\sigma$& \textbf{R - Case 1a} & \textbf{R - Case 1b} & \textbf{R - Case 1c} \\
    \midrule
    $g_{\mathrm{OO}}(r)$ & 0.02 & 6.2 & 6.1 & 6.1 \\
    $g_{\mathrm{OH}}(r)$ & 0.02 & 5.9 & 4.6 &  5.1\\
    $g_{\mathrm{HH}}(r)$ & 0.02 & 4.1 & 3.0 &  4.1\\
    $g_{\mathrm{HH}}^{\mathrm{intra}}(r)$ & 0.02 & - & - &  15.4\\
    $F_{\mathrm{D_2O}}(Q)$ & 0.0075 & 14.4 & 9.0 & 8.2 \\
    $F_{\mathrm{HDO}}(Q)$ & 0.002 & 18.0 & 18.2 & 18.7 \\
    $F_{\mathrm{'Null'}}(Q)$ & 0.00045 & 27.0 & 27.0 & 27.0 \\
    $F_{\mathrm{H_2O}}(Q)$ & 0.001 & 16.0 & 7.4 & 8.8\\
    \bottomrule
  \end{tabular}
  \caption{\label{tab:sigmaSoper} $\sigma$ parameters and R-factors for the different data sets in Case 1.}
\end{table}

\begin{table}[ht]
  \centering
  \renewcommand{\arraystretch}{1.5}
  \begin{tabular}{>{\centering\boldmath}m{0.4\columnwidth}>{\centering}m{0.2\columnwidth}>{\centering\arraybackslash}m{0.2\columnwidth}}
    \toprule
    & \boldmath{$\sigma$}& \textbf{R} \\
    \midrule
    $\Delta G_{\mathrm{D}}(r)$ & 0.000125 & 57.7\\
    $\Delta G_{\mathrm{H}}(r)$ & 0.0005 & 56.1\\
    $\Delta F_{\mathrm{D}}(Q)$ & 0.00003 & 9.6\\
    $\Delta F_{\mathrm{H}}(Q)$ & 0.00002 & 8.6\\
    \bottomrule
  \end{tabular}
  \caption{\label{tab:sigmaSalmon} $\sigma$ parameters and R-factors for the different data sets in Case 2.}
\end{table}

\section{Results and discussion}

\subsection{Case 1: H/D isotopic substitution}

Four total scattering structure factors of four mixtures of light and heavy water were considered, as seen in Fig. 2 of Ref. \cite{Soper2014}. The four compositions are: pure $\mathrm{D_2O}$, pure $\mathrm{H_2O}$, and two mixtures of light and heavy water with $x=0.5$ (denoted as 'HDO') and $x=0.64$ (denoted as 'Null')  ($x$ is the mole fraction of light water in the mixture). In the case of the 'Null' composition the hydrogen atoms have zero effective coherent scattering length and therefore, neutron scattering from this sample is sensitive to O-O correlations only. The partial radial distribution functions ($g_{\mathrm{OO}}$, $g_{\mathrm{OH}}$, and $g_{\mathrm{HH}}$) were selected from Fig 4. of Ref. \cite{Soper2014}. Simulations were carried out to fit the 7 data sets simultaneously.

The partial $g(r)$ curves were fitted only in the intermolecular regime ($r_{\mathrm{min}} =2.24$ \AA{} for $g_{\mathrm{OO}}$, $r_{\mathrm{min}}=1.29$ \AA{} for $g_{\mathrm{OH}}$, and $r_{\mathrm{min}}=1.89$ \AA{} for $g_{\mathrm{HH}}$). The total structure factors were fitted only from $Q_{\mathrm{min}} \approx 1.1$ \AA${}^{-1}$, since the low $Q$ region seemed to be the most uncertain (the $Q$ range was the same for the four TSSF-s). (The quality of the fits over the low Q regime was also checked: it was found that below $Q_{\mathrm{min}} \approx 1.1$ \AA${}^{-1}$ the curves fit poorly, so data from that region were omitted.)

Two sets of the intramolecular O-H and H-H ranges (FNC-s) were checked (see Table \ref{tab:fnc}); note that these choices influence the molecular geometry that may form. The first one is stricter, corresponding to a realistic water molecule (Case 1a), whereas the second one (yielding 'elastic' molecules, Case 1b and Case 1c) leads to a wider range of intramolecular distances, in accord with the distance ranges depicted in Fig. 4 of Ref. \cite{Soper2014}.

\begin{figure}
 \includegraphics[width=\columnwidth]{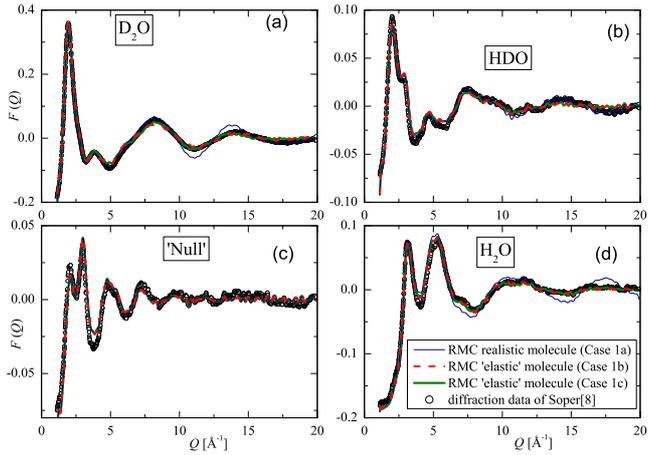}
 \caption{RMC fits to the total scattering functions of the four water samples in Case 1. (Thin solid(blue) lines: RMC results for realistic molecules (Case 1a),  dashed lines (red): RMC results for 'elastic' molecules without $g_{\mathrm{HH}}^{\mathrm{intra}}(r)$ fit (Case 1b), thick solid lines (green): RMC results for 'elastic' molecules with $g_{\mathrm{HH}}^{\mathrm{intra}}(r)$ fit (Case 1c), open circles (black) diffraction data from Fig. 2 of \cite{Soper2014})}
 \label{fig:Soperfits}
\end{figure}

\begin{figure}
\includegraphics[width=0.9\columnwidth]{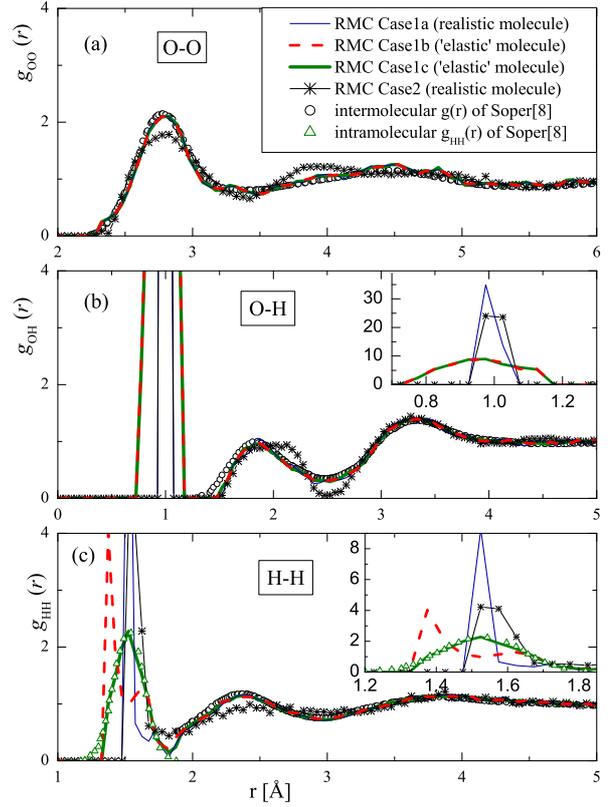}
 \caption{ Partial radial distribution functions obtained by RMC modeling. Thin solid (blue) lines: RMC results for realistic molecules (Case 1a), dashed lines (red): RMC results for 'elastic' molecules without $g_{\mathrm{HH}}^{\mathrm{intra}}(r)$ fit (Case 1b), thick solid lines (green): RMC results for 'elastic' molecules with $g_{\mathrm{HH}}^{\mathrm{intra}}(r)$ fit (Case 1c), thick (black) lines with stars: RMC results for Case 2, open circles: intermolecular $g(r)$ functions from Fig. 4. of \cite{Soper2014}, open triangles: intramolecular $g_{\mathrm{HH}}(r)$ function from Fig. 4. of \cite{Soper2014}. The insets show the intramolecular regions of the curves (b) $g_\mathrm{OH}(r)$ and (c)  $g_\mathrm{HH}(r)$. }
  \label{fig:ppcfs}
\end{figure}

The best fits obtained with realistic (as well as with 'elastic') molecules are shown in Fig. \ref{fig:Soperfits}. The goodness of fit values (R-factors) corresponding to the best fits are found in Table \ref{tab:sigmaSoper}. The PRDF-s are displayed in Fig. \ref{fig:ppcfs}. Note that the high-$Q$ parts of the $\mathrm{H_2O}$ and $\mathrm{D_2O}$ TSSF-s could not be approached sufficiently well when the 'strict' definition of water molecules was in use.

\begin{figure}
  \includegraphics[width=\columnwidth]{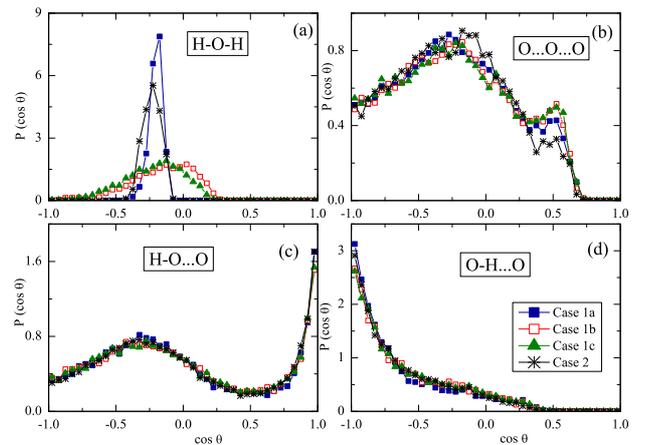}
 \caption{Distribution of the cosines of (a) H-O-H, (b) O$\cdots$O$\cdots$O, (c) H-O$\cdots$O, (d) O-H$\cdots$O angles.  (Full squares: realistic molecules in Case 1a, open squares: 'elastic' molecules without $g_{\mathrm{HH}}^{\mathrm{intra}}(r)$ fit (Case 1b), full triangle 'elastic' molecules with $g_{\mathrm{HH}}^{\mathrm{intra}}(r)$ fit (Case 1c), stars: Case 2. The lines are only guides to the eye.)} 
 \label{fig:angles} 
\end{figure}

Comparing our $g(r)$ functions with Fig. 4 of Ref. \cite{Soper2014} it can be seen that in \cite{Soper2014} the O-H and H-H intramolecular maxima are wider than those allowed by our 'strict' FNC-s. Allowing such flexibility means that the molecules must suffer elastic deformations. In any case, we repeated our RMC calculations with these more flexible intramolecular constraints, see Case 1b. Other simulation details (cutoffs, $\sigma$ parameters, etc...) were the same as previously. The TSSF-s obtained by using 'elastic' molecules are shown in Fig. \ref{fig:Soperfits}, and the corresponding R-factors are listed in Table \ref{tab:sigmaSoper}. The R-factors are smaller, i.e. the fits are better when 'elastic' molecules are applied. However, the intramolecular parts of the PRDF-s functions have become unrealistic, see Fig. \ref{fig:ppcfs}. The H-H partial radial distribution function has now two intramolecular peaks, at $r\approx 1.38$ and $r\approx 1.65$, and a local minimum at the expected distance ($r \approx 1.55$). The first peak is at the smallest value of the H-H intramolecular distance range allowed by the FNC-s (see Table \ref{tab:fnc}). The deformation of the 'elastic' molecule is also clearly visible in the distribution of the H-O-H (intramolecular) bond angles, see Fig. \ref{fig:angles} for the realistic and 'elastic' molecules. The H-O-H angle distribution of realistic molecules has a sharp peak around its most probable value (cca. 100\textdegree, $\cos \theta \approx $ -0.2). However, for the 'elastic' molecule the H-O-H angle has a broad distribution with only a hump at around 103\textdegree, while the most developed (although still small) maximum is at 90\textdegree; this pattern is rather unrealistic.  

As a further effort to improve the situation, a third simulation (Case 1c) was also performed, by fitting also the intramolecular part of $g_{\mathrm{HH}}$ from Fig 4. of \cite{Soper2014} (note that the exact origin of this part remains undisclosed in Ref. \cite{Soper2014}). That is, in this RMC calculation 8 data sets were modeled simultaneously.  The 'elastic' definition of water molecules was used, while other simulation details (cutoffs, $\sigma$ parameters) remained identical. The TSSF-s, PRDF-s and the distribution of H-O-H bond angles are presented in Figs \ref{fig:Soperfits}, \ref{fig:ppcfs} and \ref{fig:angles}. R-factors are provided in Table \ref{tab:sigmaSoper}. Clearly, the numerical values of the individual R-factors have increased somewhat, although there seem to be no visible difference in comparison with Case 1b in terms of the TSSF-s and the intermolecular parts of the PRDF-s. The intramolecular part of $g_{\mathrm{HH}}(r)$ now looks more sensible, although the 
H-O-H bond angle distribution is still too wide to be realistic.

It is instructive to notice that there are some $Q$ regions where the experimental TSSF-s could not be approached as well for the $^1\mathrm{H}$-containing samples (see particularly the 'Null' sample) as it is possible for pure $\mathrm{D_2O}$. This, together with the uncertainties of the intramolecular structure, may be considered as the manifestation of the difficult treatment of the incoherent inelastic background.

For the characterization of the local structure in liquid water, the distributions of intermolecular O$\cdots$O$\cdots$O, H-O$\cdots$O, and O-H$\cdots$O angles have also been calculated and shown in Fig. \ref{fig:angles} for the three RMC simulations in conjunction with Case 1. A more detailed discussion will be provided after results concerning Case 2 are also introduced. Here it will suffice to note that although the molecular structure for these calculations (Cases 1a-c) are rather different, this does not seem to influence structural features connected to hydrogen bonding.

\subsection{Case 2: H/D \emph{and} $^{16}\mathrm{O}/^{18}\mathrm{O}$ isotopic substitution}

The second set of data considered here is based on oxygen isotope substitution measurements \cite{Zeidler-Salmon2012, Zeidler-Salmon2011}). Four different isotopic mixtures were investigated in these publications: two heavy water ($\mathrm{D_2{ }^{18}O}$ and $\mathrm{D_2{ }^{nat}O}$) and two light water ($\mathrm{H_2{ }^{18}O}$ and $\mathrm{H_2{ }^{nat}O}$) samples. The first difference functions from Fig. 4 and Fig. 5 of Ref. \cite{Zeidler-Salmon2012} were taken for our calculations, both in $Q$ and $r$ space. These functions were calculated in the following way: after corrections the corresponding scattering cross sections were subtracted from each other, in order to eliminate the contributions from H-H or D-D correlations, together with the bulk of the inelastic scattering. The first difference functions are:

\begin{multline}
  \label{eq:deltaFD}
 \Delta F_{\mathrm{D}}(Q)=F_{\mathrm{D}}^{18}(Q)-F_{\mathrm{D}}^{\mathrm{nat}}(Q)= \\ =  c_{\mathrm{O}}^2(b^2_{{}^{18}{\mathrm{O}}}-b^2_{{}^{\mathrm{nat}}{\mathrm{O}}}) \left[ S_{\mathrm{OO}}(Q)-1 \right] + \\ + 2 c_{\mathrm{O}}c_{\mathrm{D}}b_{\mathrm{D}}(b_{{}^{18}\mathrm{O}} - b_{{}^{\mathrm{nat}}\mathrm{O}} ) \left[ S_{\mathrm{OD}}(Q) - 1 \right]
\end{multline}

and 

\begin{multline}
  \label{eq:deltaFH}
 \Delta F_{\mathrm{H}}(Q)=F_{\mathrm{H}}^{18}(Q)-F_{\mathrm{H}}^{\mathrm{nat}}(Q)= \\ = c_{\mathrm{O}}^2(b^2_{{}^{18}{\mathrm{O}}}-b^2_{{}^{\mathrm{nat}}{\mathrm{O}}}) \left[ S_{\mathrm{OO}}(Q)-1 \right] +  \\ + 2 c_{\mathrm{O}}c_{\mathrm{H}}b_{\mathrm{H}}(b_{{}^{18}\mathrm{O}} - b_{{}^{\mathrm{nat}}\mathrm{O}} ) \left[ S_{\mathrm{OH}}(Q) - 1 \right]
\end{multline}

where $c_{\alpha}$ and $b_{\alpha}$ is the atomic fraction and bound coherent neutron scattering length of chemical species $\alpha$, $S_{\alpha\beta}(Q)$ is a partial structure factor and $Q=(4\pi / \lambda ) \sin \theta$ is the magnitude of the scattering vector. Taking into account the neutron scattering lengths and the exact compositions, the weighting factors for the $[S_{\mathrm{OO}}(Q)-1]$ term in equations (\ref{eq:deltaFD}) and (\ref{eq:deltaFH}) were 0.00262 and 0.00263, and for the $[S_{\mathrm{OD}}(Q)-1]$ and $[S_{\mathrm{OH}}(Q)-1]$ terms were 0.0059 and -0.0033, respectively.

The corresponding real space functions $\Delta G_{\mathrm{D}}(r)$ and $\Delta G_{\mathrm{H}}(r)$ (which were obtained by Fourier transforming $\Delta F_{\mathrm{D}}(r)$ and $\Delta F_{\mathrm{H}}(r)$), appearing originally in Fig. 6 of \cite{Zeidler-Salmon2012}, were selected, too. These functions can be expressed from equations (\ref{eq:deltaFD}) and (\ref{eq:deltaFH}) as 

\begin{equation}
 \label{eq:deltaGD}
 \Delta G_{\mathrm{D}}(r)=0.0059 [g_{\mathrm{OD}}(r)-1] + 0.00262 [g_{\mathrm{OO}}(r)-1]
\end{equation}

and 

\begin{equation}
 \label{eq:deltaGH}
 \Delta G_{\mathrm{H}}(r)=-0.0033 [g_{\mathrm{OH}}(r)-1] + 0.00263 [g_{\mathrm{OO}}(r)-1].
\end{equation}

\begin{figure}
  \includegraphics[width=\columnwidth]{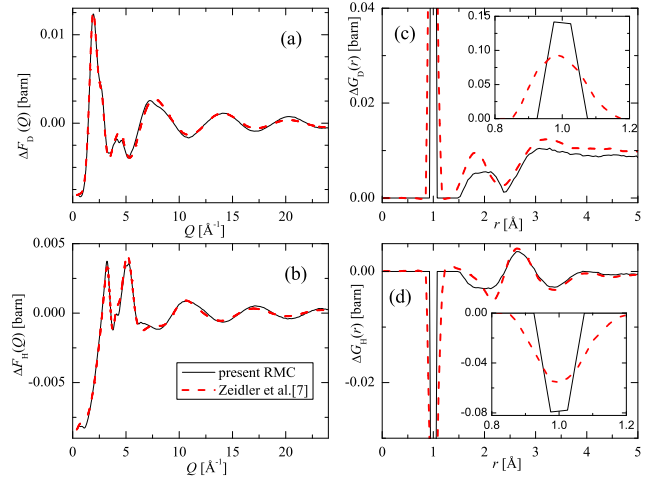}
 \caption{ RMC fits (black line) to the first difference functions of \cite{Zeidler-Salmon2012} (red dashed line).  (a) $\Delta F_{\mathrm{D}}(r)$, (b) $\Delta F_{\mathrm{H}}(r)$), (c) $\Delta G_{\mathrm{D}}(r)$, (d) $\Delta G_{\mathrm{H}}(r)$. The insets show the region of the first peak in $\Delta G_{\mathrm{D}}(r)$ (c), and the first trough in $\Delta G_{\mathrm{H}}(r)$ (d).}
\label{fig:Salmon-fit}
 \end{figure}

RMC simulations were carried out with the aim of fitting the four data sets (as mentioned above) simultaneously. Different starting configurations (random, as well as final configurations from Case 1) were used. To find the best possible agreement between RMC and experiment, simulations were carried out in three different ways: (a) At first, only the two $\Delta F(Q)$ data sets were fitted, and the two $\Delta G(r)$-s were taken into account in the second step. (b) The reverse way: $\Delta G(r)$-s were applied in first step. (c) The four data sets were fitted together from the starting configurations. The best simultaneous fits for the four functions are shown in Fig. \ref{fig:Salmon-fit}, while the corresponding partial radial distribution functions are given in Fig. \ref{fig:ppcfs}. Goodness of fit (R-factor) values are in Table \ref{tab:sigmaSalmon}. The quality of the fits of the $\Delta G (r)$ data sets clearly shows that these four data sets are not entirely consistent: the largest discrepancies show up, interestingly, in terms of the 'deuterated' $\Delta G_{\mathrm{D}}(r)$. 

Considering now the molecular structure, the situation is quite encouraging: the four target functions contain direct information on the intramolecular O-H correlations and clearly, the 'strict' FNC values were easily applicable. The distribution of intramolecular H-O-H angles (see Fig. \ref{fig:angles}) reflect far the most realistic molecular geometry: the maximum is symmetric, it is at the correct angle (above 100\textdegree), and there is no sign of any preference of either limiting value of the FNC range. (We note here that a variety of FNC-s have been tested here, too, with the hope that wider ranges would decrease differences between experimental and RMC results; as this desired outcome has not appeared, we have left the original 'strict' boundaries unchanged.)

Distributions of intermolecular O$\cdots$O$\cdots$O, H-O$\cdots$O, and O-H$\cdots$O angles are also shown in Fig. \ref{fig:angles}. Each distribution given for Case 2 is very similar to its Case 1 counterpart: the largest discrepancies can be observed for the O$\cdots$O$\cdots$O angles where the maximum corresponding to angles of about 60\textdegree (close packing, hard sphere like feature that is influenced by the nearest second neighbors) appears to be weakest for Case 2. This observation is consistent with differences between O-O PRDF-s (Fig. \ref{fig:ppcfs}) in the $r$ range of 3.5 to 5 \AA .

\section{Conclusions}

Having completed a fairly detailed Reverse Monte Carlo modeling study of the two most recent, important sets of neutron diffraction data \cite{Soper2014,Zeidler-Salmon2011}, the following findings are worth noting:

(1) Concerning the data of Soper \cite{Soper2014}, both the TSSF-s and the intermolecular parts of the PRDF-s could be modeled satisfactorily, whereas the molecular structure had to be allowed to be somewhat unrealistically too vague.

(2) Concerning the data of Zeidler et al. \cite{Zeidler-Salmon2011,Zeidler-Salmon2012}, the level consistency between $Q$ and $r$ space information was somewhat inferior, but the molecular structure appeared as more realistic.

(3) Clearly, even though the two sets of experiments (represented here by Case 1 and Case 2), as well as the two methods of data processing, could hardly be more different, essential details of the structure that concern hydrogen bonding turn out to be rather similar (in fact, nearly identical), as evidenced by Fig. \ref{fig:angles}.

(4) The above finding provides some further hints as to why the structure of liquid water still presents difficulties: the duality (large differences between experimental data sets, and close similarities between conclusions on the microscopic structure; these appear at the same time) exposed by the present study leaves the interested reader somewhat puzzled. On the other hand, we may perhaps be assured that basic features of the hydrogen bonded structure have been captured correctly over the years.

\section*{Acknowledgement}

The authors acknowledge financial support from the National Basic Research Fund (OTKA, Hungary), under grant No. K083529.

\end{document}